# Chirality transfer from chiral perovskite to molecular dopants via charge transfer states

Guan-Lin Chen[1,2], Hsinhan Tsai[1,3], Aaron Forde[4], Kai-Wei Tseng[2], Zhe-Yu Liu[2], Chi-An Dai[2], Tong Xiao[5], Mircea Coltlet[5], Leeyih Wang[2], Sergei Tretiak[4], Wanyi Nie[1]*

1. Department of Physics, SUNY University at Buffalo, Buffalo, NY, USA

2. Center for Condensed Matter Sciences, National Taiwan University, Taipei, 10617, Taiwan

3. Department of Chemistry, University of California, Berkeley, CA, USA

4. Theoretical Division, Los Alamos National Laboratory, Los Alamos, NM, USA

5. Center for Functional Nanomaterials, Brookhaven National Laboratory, Upton, NY, USA

*Corresponding to: wanyinie@buffalo.edu

**Abstract**

Chiral perovskites are emerging semiconducting materials with broken symmetry that can selectively absorb and emit circularly polarized light. However, most of the chiral perovskites are typically low-dimensional structures with limited electrical conductivity and their light absorption occurs in the UV region. In this work, we find doping 2,3,5,6-Tetrafluoro-7,7,8,8-tetracyanoquinodimethane (F4TCNQ) in the chiral perovskite matrix can improve the electrical conductivity with an addition of visible light absorption through the emerging charge-transfer electronic states. The new absorption feature exhibits strong circular dichroism adapted from the chiral matrix, which is indicative of a chirality transfer from the host to the guest via an electronic coupling. The charge transfer state is validated by transient absorption spectroscopy and theory modeling. Quantum-chemical modeling identifies a strong wave function overlap between an electron and a hole of the guest-host in a closely packed crystal configuration forming the charge-transfer absorption state. We then integrate the doped chiral perovskite film in photodetectors and demonstrate a selective detection of circularly polarized light both in the UV and visible range. Our results suggest a universal approach of introducing visible photo absorption states to the chiral matrix to broaden the optical active range and enhance the conductivity.

**Introduction**

Chirality is an important symmetry property of molecular systems [1] : A chiral molecule cannot be superimposed on its mirror image, which breaks the molecule's inversion symmetry [2]. Many biologically active molecules are chiral, including such important systems as amino acids or enzymes. Nature broadly employs chiral systems, for instance, for biological specificity, molecular recognition and asymmetric synthesis [3-7]. Inspired by this Nature's selection approach, semiconductors crafted in chiral structures can potentially enable selectivity to opto-electronic signals. The electric dipoles of the excited states in chiral semiconductors are arranged along the helicity of the crystal structures with anisotropic absorbance and emission of circularly polarized light [8,9]. Additionally, chiral-induced spin selectivity effect [10,11] is proposed in the chiral semiconductors which can discriminatorily transport spin-polarized carriers caused by a strong spin-orbital coupling effect owing to the broken symmetry.

Chiral perovskites are one of such chiral semiconductor systems that have shown great promise in chiral sensing applications. In a chiral perovskite structure family, chiral organic molecules incorporated into lead-halide octahedrons ($PbX_6$) drive rotations of the $PbX_6$ cages following the chirality of the molecules. The dipole moment of the excited states thus orients with the helix $PbX_6$ rotation direction to impose optical selectivity to circular polarization state of photons [9]. With strong circular dichroism (CD) demonstrated [12-17] chiral perovskites were integrated into photodetectors for the direct detection of circularly polarized light. The detectors were capable of differentiate the handness of the incoming circularly polarized light by its distinct change in photoconductivity, delivering anisotropic factors of 0.1 ~ 0.2 [17-21]. Circularly polarized light emission from the chiral perovskites was also possible [18,22,23], which are promising photonic sources with polarized states. Utilizing the chiral induced spin selectivity, a chiral perovskite thin film was employed for polarized carrier injection to quantum dots or III-V light emitter for circularly polarized electroluminescence, with a polarization up to 15% [24-26]. While these pioneer demonstrations open new avenues for chiral opto-electronics, the organics in the chiral perovskites were considered insulators that add barriers to the electrical conduction of the charge carriers [27-29]. Chiral perovskites are typically low-dimensional structures [8,30,31] that underpin quantum confinement. Consequently, chiral perovskites have a wide optical band gap and limited responsivity to visible light. Due to their low conductivity and lack of photo-activity in the visible range, low dimensional chiral perovskite materials face significant limitations in their potential electronic applications.

To bridge this gap, here we introduce a universal approach that employ molecular dopant into the chiral perovskite matrix to enhance the electrical conductivity and extend absorption to the visible range. By mixing 2,3,5,6-Tetrafluoro-7,7,8,8-tetracyanoquinodimethane (F4TCNQ) molecules into the chiral perovskite matrix, we observe the emergence of an optically active charge transfer state formed between the chiral perovskites host and the F4TCNQ guest, which adds a visible absorption band in the 500-700 nm wavelength range. Interestingly, both the intrinsic absorption band from the perovskite matrix and the charge transfer absorption feature exhibit strong circular dichroism, indicative of a chirality transfer from the host to the guest. Transient absorption spectroscopy confirmed the presence of charge transfer states at the interface. Furthermore, atomistic time-dependent density functional theory (TDDFT) simulations of the chiral perovskite/F4TCNQ interface corroborate the experimental observations. Specifically, our calculations suggest that substitutional doping of F4TCNQ enhances electronic coupling leading to the formation of optically bright charge-transfer states. Finally, we demonstrate the direct detection of circularly polarized light by integrating our host-guest structure in a photo-detector. Due to the chirality transfer, our detector can discern the chirality of photon across both the blue and visible spectral ranges. Altogether, our findings suggest a general approach for chirality transfer from the host matrix to the guest dopant via an electronically coupled charge transfer state.

**Results**

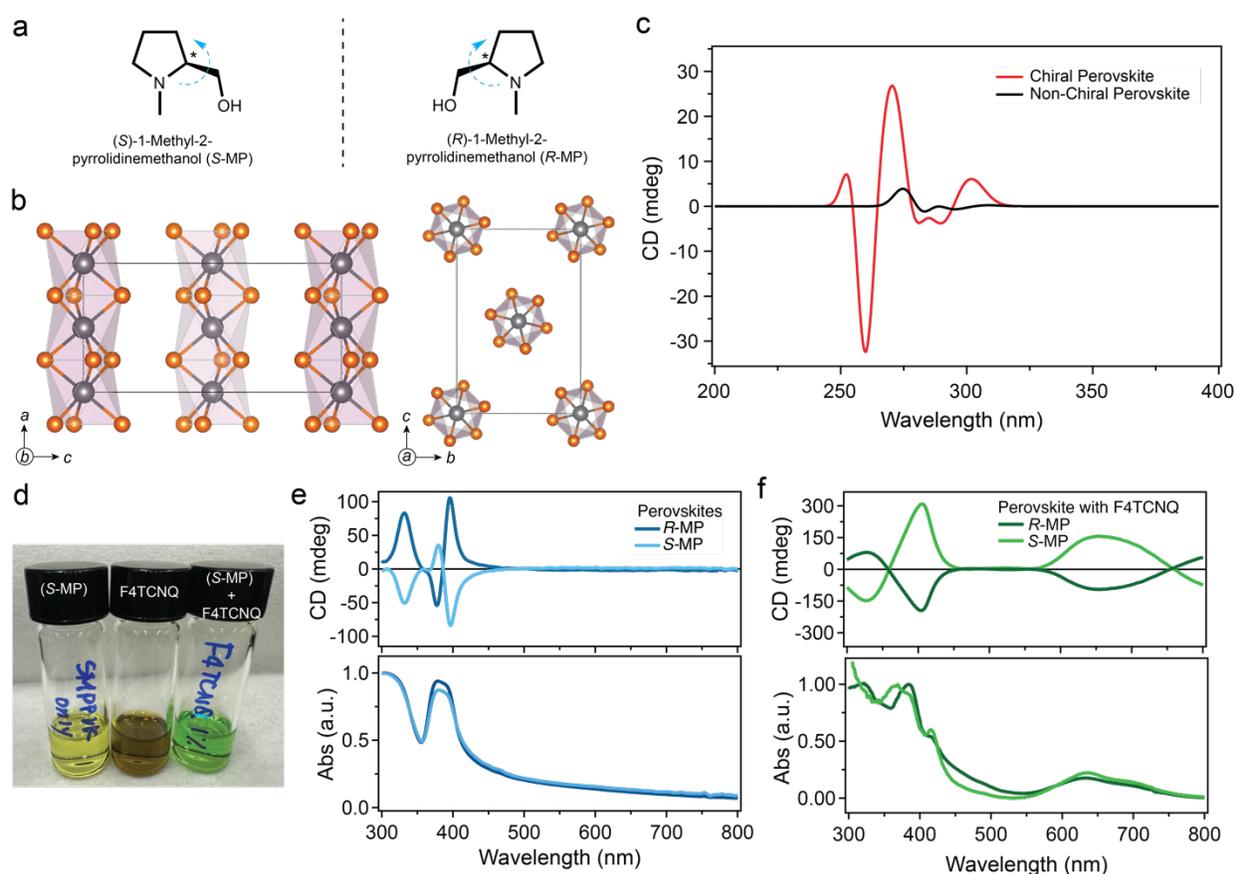

**Figure 1** (**a**) Chiral molecules (*S*-MP and *R*-MP) used in this study. (**b**) Crystal structure of the chiral perovskite structures as viewed down from the *b* axis and *a* axis, respectively. (**c**) Calculated CD spectrum for the chiral perovskites and non-chiral perovskite structure using the reported structures. (**d**) Photographs of the *S*-MP-perovskite (left, light yellow), F4TCNQ (middle. brown) and *S*-MP-perovskite with 10% F4TCNQ dopant (right, bright green). (**e, f**) CD and absorption spectra measured for pristine chiral perovskite thin films (**e**) and F4TCNQ doped chiral perovskite thin films (**f**).

We start with synthesizing chiral perovskite structure using chiral enantiomer (($S$)-1-Methyl-2-pyrrolidinemethanol ($S$-MP) and ($R$)-1-Methyl-2-pyrrolidinemethanol ($R$-MP)) as shown in Fig.1a. The material growth procedures are detailed in Supplementary Method Section. Following material synthesis and recrystallization, we obtained needle-like palm-yellow single crystals (Supplementary Fig. S1), indicative of a 1D structure formation. We performed X-ray scattering on the single crystal and determined the crystal structure as illustrated in Fig. 1b. The crystal structure exhibits one-dimensional face-sharing octahedra with Sohncke type space group of $P2_12_12$, an analogue of the structure of $((CH_3)_3S^+)_4Pb_3Br_{10}$ [32]. Using this crystal structure, we designed model simulations to gain atomistic insights into the chirality property of the obtained crystals. Our approach addresses two fundamental aspects: the chirality of the $PbI_6$ sublattice[33] and chirality transfer via electrostatic interactions with the chiral molecules [34,35]. Specifically, we use the X-ray resolved crystallographic structure of the $PbI_6$ sublattice, crystalized with chiral MP molecules, to compute the circular dichroism (CD) spectrum using TDDFT methodology, as briefly outlined in the Supplementary Methods Section. As a control, we also calculate the CD spectrum for the perovskite synthesized with the achiral MP molecule using a published crystal structure file (Supplementary Figure S2) [36]. Figure 1b compares the computed CD spectra. The plot reveals that the perovskite synthesized with chiral MP exhibits much enhanced CD intensities whereas the structure synthesized with achiral MP shows negligible signal. This result supports the concept of chirality transfer from a chiral molecule to the 1D crystal structure of the perovskite. In our case, a structural chiral distortion in the $PbI_6$ 1D chain is identified as a dominating mechanism. We also modeled CD spectrum when chiral MP are placed on the surface the perovskite lattice without driving the $PbI_6$ cage rotations. If the chirality transfer is mediated by electrostatic interactions we would expect to see mirrored CD spectrum when comparing models containing the MP enantiomers. We find that their CD spectrum are nearly identical and only slightly deviate from the 'bare' perovskite (Supplementary Fig. S2). This comparison further confirms that the CD originates from the chirality imprinted in the PbI$_6$ inorganic lattice driven by the chiral MP molecules.

We next experimentally measured the CD and absorption spectra of the perovskite thin films, as shown in Fig. 1(d-e) under illumination with left-handed and right-handed circularly polarized light. To eliminate the influence of the linear dichroism, we have tested the CD spectra after rotating the samples at various angles with respect to the incident light. The results are shown in Supplementary Figs. S4-6. Next, F4TCNQ was added to the precursor, with the weight percentage varied from 0.1% to 10%. Figure 1d is the photograph of the obtained solutions. With F4TCNQ, the solution turns into a bright green color, different from either pristine solution. We spin cast thin films with these solutions and then annealed them at 100 ºC for 10 minutes. Like the solution, the pristine chiral perovskite thin film has a transparent yellow color, whereas the doped film exhibits a green color (see film photo shown in Supplementary Fig. S1). Figure 1e-f shows the CD and absorption spectra of the pristine and doped *S*-MP and *R*-MP chiral perovskite thin film. The pristine films exhibit two "derivative shaped" CD features crossing zero at 360 nm and 390 nm, corresponding to the two absorption peaks of the perovskite material. Near the absorption peak, the *R*-MP (*S*-MP) sample exhibits a negative (positive) CD at higher energy and a positive (negative) feature at lower energy, manifests a Cotton effect [37,38]. Figure 1f shows the CD spectrum of the F4TCNQ doped chiral perovskites thin film. Here CD signals emerge near the absorption peaks of the perovskite band edge and broaden towards 416, along with a broad visible light absorption emerging between 550 nm and 750 nm, which contributes to the green color. F4TCNQ dopant is known for its strong electron-accepting ability due to its electron-deficient nature. We speculate that the new absorption feature can be attributed to the charge-transfer state formed at the donor-acceptor interface, as documented by prior reports on organic systems doped with F4TCNQ [39-41]. The fact that CD signals are also detected near the broad peak in the 550 nm to 750 nm range suggests the hypothesized charge-transfer states also inherits the chiroptical properties from the perovskite.

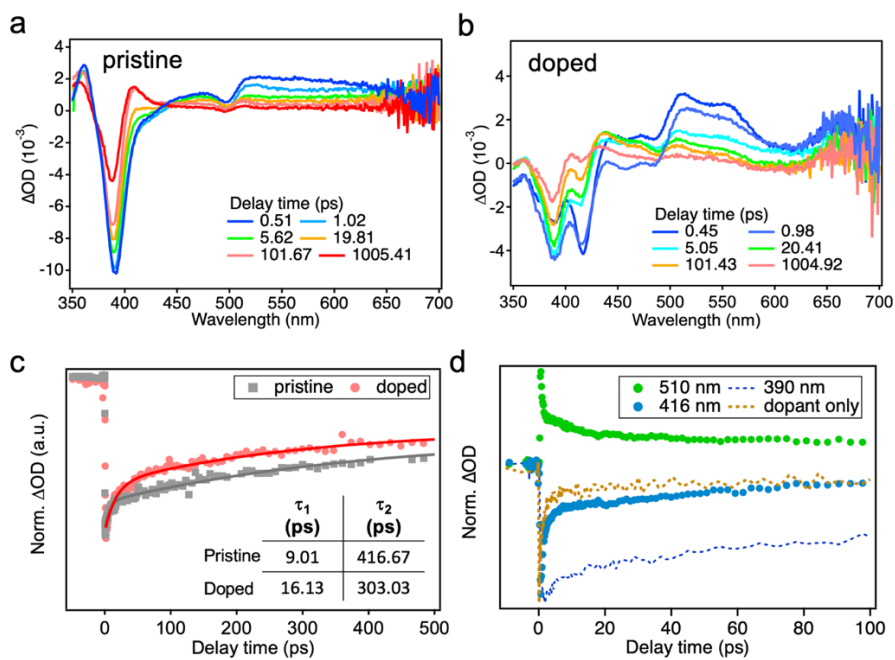

**Figure 2** Broadband transient absorbance vs wavelength obtained using 290nm pump excitation for **(a)** a pristine perovskite and **(b)** a 10 % F4TCNQ doped perovskite thin films at various delay times following the pump excitation. **(c)** The 390 nm ground state bleaching peak intensity as a function of delay time after pump of pristine sample and doped sample. The extracted time constants from the double exponential fitting are summarized in the same plot. **(d)** Peak intensity vs delay time plot comparing peaks at 416 nm and 510 nm of the doped sample. The peak at 390 nm of the doped sample, and the ground state bleaching signal of the dopant itself (when pumped by 340 nm laser) are also plotted for comparison. The kinetic time constants associated with various peaks of these spectra are presented in Supplementary Table S1.

To elucidate the electronic transitions associated with the different absorption features, we performed femtosecond pump-probe broadband transient absorption spectroscopy (TA) on both pristine and doped thin-film samples. From the spectra and kinetic analysis of each peak in the TA spectrum, we confirm the presence of charge-transfer coupling between the chiral perovskite and the F4TCNQ dopant.

The TA spectra, probed at various delay times following the pump pulse for both samples, are plotted in Fig. 2a-b. We used a 290 nm laser pump for excitation and a broadband source to probe the transient absorption in the UV-Vis spectral region. We specifically selected a 290 nm pump laser for further analysis, which solely excites the chiral perovskite matrix while avoiding direct excitation of the dopant. Therefore, any signal associated with the dopant would result from electronic transitions coupled directly to the chiral perovskite host. In the TA spectrum of the undoped chiral perovskite displayed in Fig. 2a, a pronounced negative feature emerges at 390 nm within sub-picosecond time range after pumping, which is attributed to the ground state bleaching[42]. A positive feature near 360 nm and a weaker positive peak near 490 nm are also observed in this sample, corresponding to photo-induced absorption transition in the high and low energy bands. In the doped sample, the same ground state bleaching feature at approximately 390 nm is observed, along with an additional negative peak at 416 nm, as shown in Fig. 2b. This region corresponds to where F4TCNQ absorbs, as indicated by the steady state absorption spectrum (see Supplementary Fig. S3). We found that F4TCNQ dopant cannot be directly excited by the 290 nm laser pump used in this experiment. While pumping with 340 nm is possible, it yields a different ground state bleaching feature of F4TCNQ only sample at around 600 nm (see Supplementary Figs. S8-9). Therefore, the ground state bleaching at 416 nm in the doped sample originates from the formation of the charge transfer state. Moreover, a pronounced positive photo-induced absorption peak is observed in the doped sample between 500 nm and 600 nm. While this peak overlaps with a weak, broad absorption band in the pristine sample probed by TA, we attribute it to an electronic transition to a different charge transfer excited state, due to its distinct vibronic shape and different recombination kinetics, which will be discussed in the next paragraph.

We further analyzed the kinetics of the ground state bleaching and photo-induced absorption features for both samples, as shown in Fig. 2c-d, to determine the origin of each peak. In Fig. 2c, we overlay the kinetic curves of the 390 nm peak for both the pristine and doped samples. Noticeably, the decay curve of the pristine perovskite sample can be fitted by a double exponential function, suggesting two predominant recombination pathways for the excited state relaxation to the ground state. The pristine sample has a fast component with a time constant ($\tau_1$) of 9.01 ps followed by a slow decay with a time constant ($\tau_2$) of 416.67 ps. The decay curve of the doped sample can be fitted with the same function, yielding $\tau_1$ = 16.13 ps and $\tau_2$ = 303.03 ps for the fast and slow components, respectively. The presence of the dopant in the perovskite matrix noticeably extends the short time constant while shortening the long component. The short component $\tau_1$ is typically assigned to exciton diffusion[43,44]. Thus, in presence of the charge-transfer state, the exciton diffusion lifetime is extended due to charge separation. The long component $\tau_2$ is often attributed to charge recombination. The reduction in $\tau_2$ in the doped sample can be understood as accelerated electron injection back into the perovskite's valence band, which return the system to the ground state more quickly[43].

In Fig. 2d, we compare the kinetics of the new negative peak at 416 nm and the new positive peak at 510 nm with the ground state bleaching peak at 390 nm from the perovskite. The kinetics of the F4TCNQ's ground state bleaching peak, extracted from Supplementary Figs. S8-9, are plotted in the same graph, when pumped by a 340 nm laser. All curves can be fitted with two decay constants, as detailed in Supplementary Table S1. Both 416 nm and 510 nm peaks have $\tau_1$ of 1.32 ps and 4.38 ps, respectively, followed by $\tau_2$ of 131.58 ps and 196.08 ps. In stark contrast, the fitted $\tau_1$ and $\tau_2$ of the F4TCNQ-only sample are extremely short, at 0.19 ps and 1.53 ps, respectively. The slower decay behavior of the doped sample replicates the kinetics of the perovskite host matrix, with lifetimes significantly shortened in presence of the dopant. These results suggest that the new features at 416 nm and 510 nm do not represent the electronic transitions within the F4TCNQ molecule itself but are associated with charge transfer and recombination processes at the interface.

From the TA data analysis, we conclude the charge transfer kinetics as the following. The chiral perovskite matrix is first pumped, and charge rapidly transfer from the excited state of the chiral matrix to the dopant at fast rate, with $1/k_{CT}$ <1 ps. The new ground state bleaching and photo induced absorption features correspond to electronic transitions to the charge-transfer state. This indicates a strong interaction between the donor (chiral perovskite, host) and the acceptor (F4TCNQ, guest) components.

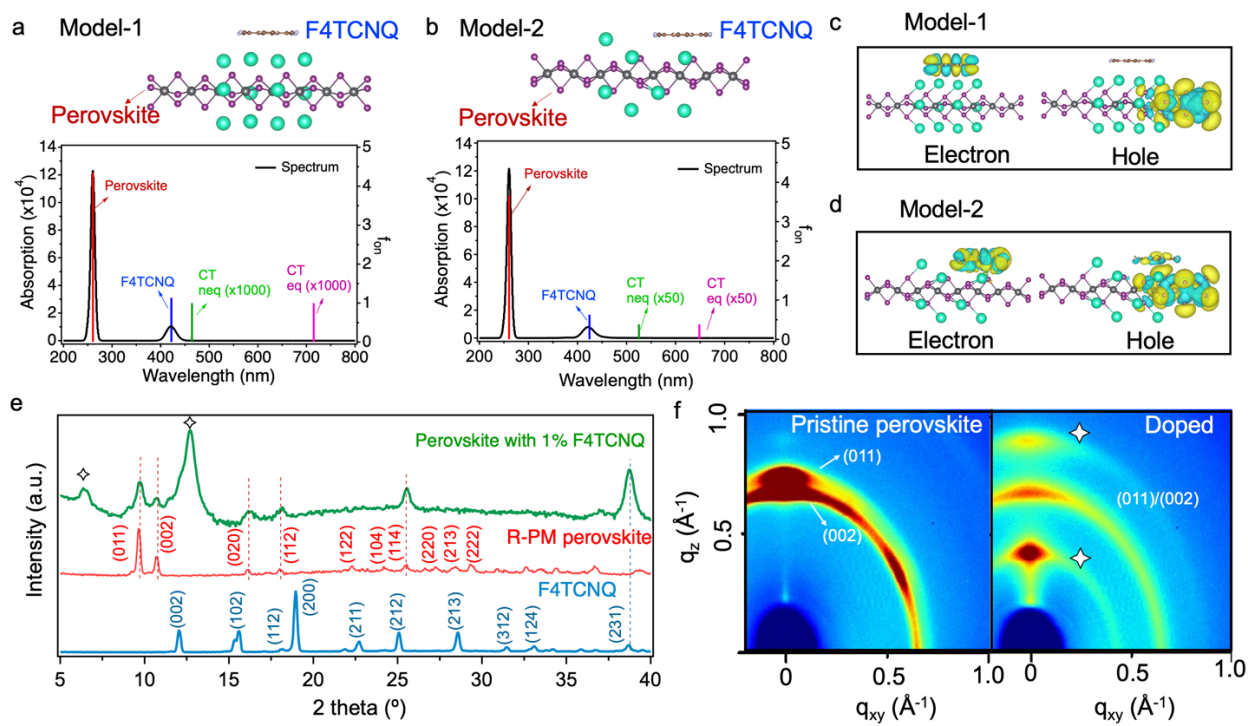

**Figure 3 Electronic structure modeling and crystalline structure analysis.** The top schematics show model-1 and model-2 structural motifs. Simulated absorption spectrum (black curve) and oscillator strengths for perovskites (red stick), F4TCNQ (blue stick), as well as composed system in linear-response equilibrium (pink) and state-specific non-equilibrium (green) methods based on model-1 (**a**) and model-2 (**b**). Note that the oscillator strengths in the composed system in (**a**) and (**b**) are magnified by 1000 and 50 times to be visible in the graphs. (**c,d**) Calculated natural transition orbitals of the charge-transfer state computed with state-specific non-equilibrium method for models-1 and model-2, respectively. (**e**) Powder X-ray diffraction pattern of F4TCNQ, chiral perovskite and F4TCNQ doped chiral perovskite with indexes labeled. (**f**) Gracing incidence wide-angle X-ray scattering (GIWAXS) of both samples zoomed into the low-angle range.

To gain insight into electronic features at the host-guest interface, we next performed quantum chemical calculations (see Methods) using two model structural arrangements of the guest in the host matrix: In model-1, F4TCNQ is intercalated between the perovskite chains, while in model-2, F4TCNQ substitutionally replaces the 'A' cation (*R/S* MP cations) in the perovskite structure, as illustrated in the top panels of Fig. 3(a,b), respectively. We computed the electronic structure of these systems, focusing on the appearance of charge transfer excitations. Crucially, we account for the dielectric effects of the perovskite lattice using state-specific solvation models, which are known to significantly influence the transition energies of charge-transfer states, as detailed in Methods section. The resulting simulated absorption spectra (black curve) along with the contributing oscillator strengths ($f_{ON}$) of the two model composite systems, as well as isolated perovskite and F4TCNQ molecule, are plotted in the same graphs. Notably, Figure 3(a,b) displays simulations results from both state-specific non-equilibrium (CT, non-eq) and equilibrium (CT, eq) solvation. The charge transfer peak in the computed absorption spectra exhibits a significant red-shift relative to the excitonic absorption at the perovskite band edge for both models (particularly for equilibrium solvation), in-line with experimental observations. Figure 3(c,d) displays the natural transition orbitals (NTOs) of the lowest excited state computed for models 1 and 2. NTOs decompose a given electronic excitation into its constituent electron and hole orbitals [45]. We observe distinct spatial localization of electron (on the molecule) and hole (on the perovskite) orbitals in both models. This highlights that upon doping, the electron transfers from the perovskite to F4TCNQ owing the election deficiency of the latter. In the ground state our simulations indicate some partial electron transfer (~0.1e) supported by improved electrical conductivity measurements discussed later. Upon optical excitation, more electronic density is getting transferred to F4TCNQ (~1e) depending on a particular model structure. This confirms our assignment of the red-shifted optical features to charge-transfer states introduced by the dopant molecule.

Importantly, the degree of electron transfer and the resulting modifications of the absorption spectra significantly depend on the relative alignment of the perovskite and F4TCNQ. In model-2, when the F4TCNQ dopant substitutes for A-site cations, the hole wavefunction becomes spatially delocalized between the perovskite and the molecule, resulting in a notable overlap with the electron wavefunction. Consequently, we observe a 20 fold enhancement in oscillator strength in Fig. 3b, whereas in model-1, where the dopant is intercalated, the charge-transfer state is essentially optically dark, as shown in Fig. 3a. This can be explained by the enhanced electronic coupling when the dopant is closer to the perovskite lattice leading to improved overlap of the electron and hole wavefunctions (Fig. 3(c,d)).

Thus, our quantum-chemical modeling suggests that a closely packed host-guest crystal structure will yield a bright charge transfer state with non-vanishing oscillation strength. Experimentally, we conducted structure characterization to probe the arrangement of the guest-host thin film structures, with the results shown in Fig. 3(e-f). Figure 3(e) displays the X-ray diffraction (XRD) patterns for F4TCNQ (blue), undoped chiral perovskites (red), and the host-guest thin film (green). The XRD patterns for F4TCNQ and chiral perovskites align well with the diffraction patterns of orthorhombic $P$bca and monoclinic space group $P2_1/c$, respectively, calculated from their CIF files extracted from the X-ray scattering. In the doped perovskite sample, several peaks correspond to those of the pristine chiral perovskite XRD peaks, indicating the presence of the host crystalline structure. Additionally, two new peaks appear at the lower angle range that do not match either the host or guest crystal structures. These diffraction peaks correspond to a long-range ordering of the crystalline structure. To examine the low angle peak more closely, we measured Gracing incidence wide angle X-ray scattering (GIWAXS) maps to further confirm the new features (Fig. 3f). The full images of $R$-MP and $S$-MP chiral perovskites, along with those with dopants, are shown in Supplementary Fig. S12. In the pristine film, we identify the main (011) and (020) peaks at 0.68 Å$^{-1}$ and 0.73 Å$^{-1}$, respectively. Upon doping, both peaks are weakened and broadened (see GIWAXS linecut in Supplementary Fig. S13), indicating that the dopant introduces disorder into the perovskite matrix. Consistently with the PXRD patterns, two new peaks emerge at 0.46 Å$^{-1}$ and 0.9 Å$^{-1}$, corresponding to a d-spacing of 13.78 Å and 6.98 Å, respectively. The ordered structures with d-spacings of 13.78 Å and 6.98 Å likely correspond to two F4TCNQ molecules (d = 7.3 nm) stacked with intercalation [46] between two perovskite chains. The F4TCNQ molecules gradually fill the space between neighboring perovskite backbones, weakening the perovskite's crystalline packing. It worth noting that the two new peaks in the doped samples are predominately oriented along the out-of-plane directions, with an intense scattering signal along the $q_z$ direction. From the X-ray scattering analysis, we can conclude that F4TCNQ molecules do not reside outside of the crystalline perovskite domains, but rather diffuse into the lattice, forming a closely packed super molecular structure. This structure aligns with the scenario described in model-2, where F4TCNQ is in a close contact with the $PbI_6$ backbone, thereby introducing optically active charge transfer states.

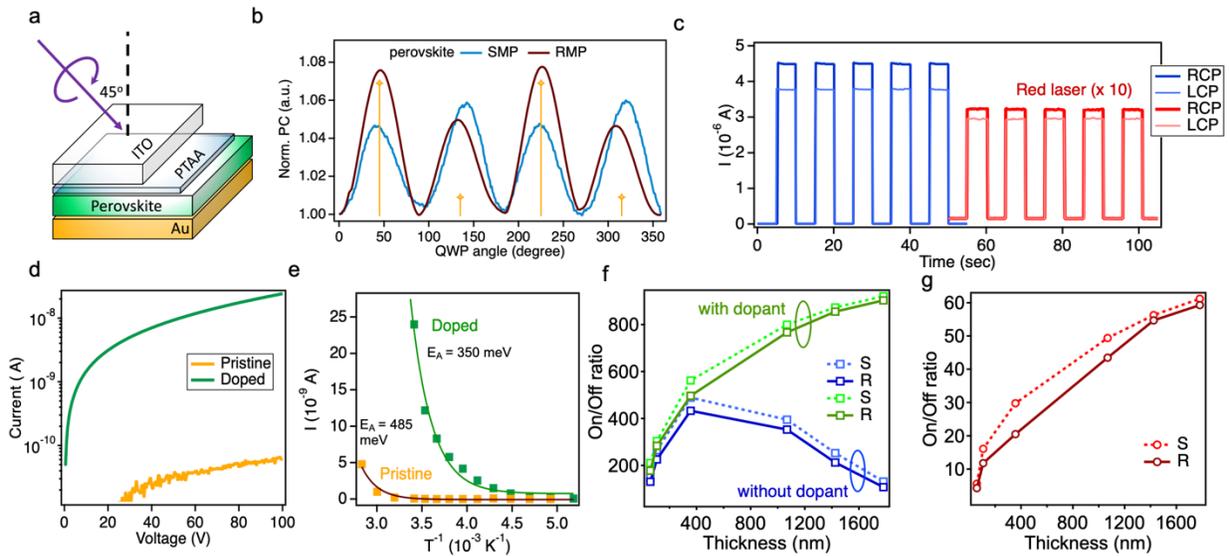

**Figure 4 Photo detector for polarized light detection. (a),** schematic illustration of the photo-detector and our measurement configuration. **(b),** photocurrent as a function of quarter wave plane rotation angle with respect to the linear polarizer direction. **(c),** time evolution of the photocurrent of *R*-MP-chiral perovskite detector when illuminated by right-handed and left-handed circularly polarized lasers, laser wavelengths are 405 nm and 635 nm. **(d),** dark current-voltage curves of chiral perovskite with and without dopant measured in dark. **(e),** temperature dependent dark current plotted against 1/T. **(f),** extracted photocurrent on/off ratio of chiral perovskite device at various absorber thicknesses when illuminated by a 405 nm laser source. **(g),** extracted photocurrent on/off ratio when illuminated by the red laser.

Finally, we demonstrate a photodetector made with the guest-host chiral perovskite thin film for the direct detection of circularly polarized light. Figure 4(a) illustrates the device structure used for the photodetector demonstration in this work. Details on the device optimization and results are presented in the supplementary information. The incoming light was maintained at 45-degree angle with respect to normal incidence to eliminate the preferential scattering effect of linearly polarized light by the crystalline domains. Figure 4b shows the photocurrent changes with quarter wave plate rotation angle respect to the linear polarizer's direction. The device made with *R*-MP-perovskite absorber shows stronger photocurrents when the incoming light is right-handed circularly polarized (quarter wave plate is at 45º, 135º, 225º and 315º to the linear polarizer's axis), whereas the photocurrent produced by the left-handed circularly polarized light is weaker. In contrast, the perovskite made with *S*-MP ligands shows the opposite trend producing lower photocurrent in response to right-handed circularly polarized light. Figure 4c shows the photocurrent response of our *R*-MP perovskite detector when illuminated by circularly polarized blue and red light sources (635 nm), with multiple on/off cycles. The photocurrent is consistently enhanced across multiple cycles when the detector is illuminated by right-handed circularly polarized light, suggesting an anisotropic detectivity. Interestingly, anisotropic detectivity is observed both blue and red light, which excite the intrinsic peak and charge transfer state respectively. The anisotropic factor (g) is calculated to be 0.18 and 0.12 for blue and red-light sources, respectively.

To understand the effect of the dopant on the material's conductivity change, we measured the temperature dependent current voltage curves of both devices in the dark. The results are shown in Fig. 4(d-e). We observe that the room temperature dark conductivity of the doped film is improved by more than 2 orders of magnitude comparing to that of without dopant. Figure 4e displays the extracted dark current against 1/T. In both devices, the dark current grows exponentially with temperature, suggesting a thermally activated transport mechanism attributed to the charge delocalization across conducting sites. The thermal activation energy can determined by the Arrhenius equation. The device with 1% F4TCNQ dopant has a lower activation energy ($E_A$ = 350 meV) compared to the undoped device ($E_A$ = 480 meV). The F4TCNQ creates a conducting channel that facilitates electron hopping the through the chiral perovskite matrix, thereby lowering the energy barrier. With a lower energy barrier for carrier conduction, there is a higher chance of extracting the charge carriers from the material. Next, we measured the photo-to-dark current ratio (on/off ratio) of devices with varying film thicknesses. The results are shown in Fig. 4(f-g). When illuminated by a blue light source, the undoped device shows a rapid raise in its on/off ratio when the film thickness increases from 20 nm to 70 nm, due to improved photo absorption. However, the on/off ratio begins to decline with further increases in thicknesses owing to poor conductivity. In contrast, the on/off ratio of the doped device continuously increases with thickness and levels off around 200 nm, aided by the improved conductivity. Figure 4(g) illustrates the on/off ratio when illuminated by a red laser, which excites the new charge transfer absorption feature. This trend mirrors that observed with the blue laser: the on/off signal increases with thickness benefitted from the higher electrical conductivity from doping.

**Conclusion:**

Our joined experimental and theory analysis clearly demonstrates that doping chiral perovskites with F4TCNQ molecules introduces a charge-transfer states, which adds visible light absorption ability. Remarkably, these new emission features inherit chiral properties from the parent system. We attribute observed the chirality transfer to two possible origins. First, it may be caused by electron transfer between the donor and acceptor. According to our model-2, the electron and hole orbitals are localized at the perovskite chain and F4TCNQ, respectively, forming the charge transfer excited state. Since the transition dipoles of the excited states align with the chiral arrangement of the perovskite lattice due to the distribution of the hole orbital, the resulting charge-transfer states also become sensitive to circularly polarized light. Second, chirality can be transferred through crystal packing between the donor and acceptor. From our X-ray scattering analysis, F4TCNQ molecules intercalate between the perovskite chains, forming a superlattice. In this case, chiral perovskite serves as a template, allowing F4TCNQ to grow with its chiral structure. As a result, the new charge transfer absorbing states adopt the chiral arrangement, delivering noticeable circular dichroism at the visible absorption range. Altogether, our work addresses the lack of visible light response in low dimensional chiral perovskites through molecular doping. The chirality transfer via electronic coupling between guest and host structure represents a general mechanism applicable to other chiral semiconductors. Furthermore, molecular doping in low dimensional chiral perovskites offers practical strategies to improve electrical conductivity, creating opportunities for photodetectors sensitive to circularly polarized light and other optoelectronic applications.

**Acknowledgement**

W.N. acknowledge the start-up funding support from the New York research foundation of SUNY-Buffalo. Part of the research used resources of the Center for Functional Nanomaterials (CFN), which is a U.S. Department of Energy Office of Science User Facility, at Brookhaven National Laboratory under Contract No. DE-SC0012704. The work was performed in part at the Center for